%% file: main.tex
\documentclass[runningheads]{llncs}
\usepackage{adjustbox}
\usepackage[utf8]{inputenc} %
\usepackage[T1]{fontenc}    %
\usepackage{hyperref}       %
\usepackage{url}            %
\usepackage{booktabs}       %
\usepackage{amsfonts}       %
\usepackage{nicefrac}       %
\usepackage{microtype}      %
\usepackage{xcolor}         %
\usepackage{marvosym}
\usepackage{xspace}
\usepackage{xcolor}
\usepackage{soul}
\usepackage[framemethod=tikz]{mdframed}
\usepackage{enumitem}
\usepackage{wrapfig}
\usepackage{caption}
\usepackage{multirow}
\usepackage{subfigure}

\newcommand{\ie}{\emph{i.e.,}\xspace}
\newcommand{\eg}{\emph{e.g.,}\xspace}

\newcommand{\paratitle}[1]{\vspace{1.5ex}\noindent\textbf{#1}}

\newcommand{\ignore}[1]{}

\definecolor{condition}{HTML}{E0EBF6}
\DeclareRobustCommand{\condition}[1]{\sethlcolor{condition}{\hl{~#1~}}}
\definecolor{candidate}{HTML}{E5EFDB}
\DeclareRobustCommand{\candidate}[1]{\sethlcolor{candidate}{\hl{~#1~}}}

\definecolor{ginger}{rgb}{0.69, 0.4, 0.0}
\definecolor{pastelorange}{rgb}{1.0, 0.7, 0.28}
\definecolor{teal}{rgb}{0.0, 0.5, 0.5}
\definecolor{thistle}{rgb}{0.85, 0.75, 0.85}

\begin{document}
\title{Large Language Models are Zero-Shot Rankers for Recommender Systems}

\renewcommand\footnotemark{}

\author{Yupeng Hou\inst{1,2}\textsuperscript{$\dagger$} \and
Junjie Zhang\inst{1}\textsuperscript{$\dagger$} \and
Zihan Lin\inst{3} \and
Hongyu Lu\inst{4} \and
Ruobing Xie\inst{4}, \\
Julian McAuley\inst{2} \and
Wayne Xin Zhao\inst{1}\textsuperscript{\Letter}
\thanks{$\dagger$ Equal contribution.}
  \thanks{\Letter\ Corresponding author.}}
\authorrunning{Hou et al.}

\institute{Gaoling School of Artificial Intelligence, Renmin University of China \and
UC San Diego \and
School of Information, Renmin University of China \and
WeChat, Tencent\\
\email{yphou@ucsd.edu\quad junjie.zhang@ruc.edu.cn\quad batmanfly@gmail.com}}
\maketitle              %
\begin{abstract}
    Recently, large language models~(LLMs) (\eg GPT-4) have demonstrated impressive general-purpose task-solving abilities, including the potential to approach recommendation tasks.
    Along this line of research, this work aims to investigate the capacity of LLMs that act as the ranking model for recommender systems.
    We first formalize the recommendation problem as a conditional ranking task, considering sequential interaction histories as \emph{conditions} and the items retrieved by other candidate generation models as \emph{candidates}.
    To solve the ranking task by LLMs, we carefully design the prompting  template
    and conduct extensive experiments on two widely-used datasets.
    We show that LLMs have promising zero-shot ranking abilities but (1) struggle to perceive the order of historical interactions, and (2) can be biased by popularity or item positions in the prompts. We demonstrate that these issues can be alleviated using specially designed prompting and bootstrapping strategies. Equipped with these insights, zero-shot LLMs can even challenge conventional recommendation models when ranking candidates are retrieved by multiple candidate generators.
    The code and processed datasets are available at \textcolor{blue}{\url{https://github.com/RUCAIBox/LLMRank}}.

\end{abstract}

\input{sec-1-intro}
\input{sec-2-method}
\input{sec-3-experiment}

\input{sec-4-related-work}

\section{Conclusion}\label{sec:conclusion}

In this work, we investigated the capacities of LLMs that act as the zero-shot ranking model for recommender systems. 
To rank with LLMs, we constructed natural language prompts that contain historical interactions, candidates, and instruction templates. We then propose several specially designed prompting strategies to trigger the ability of LLMs to perceive orders of sequential behaviors. We also introduce bootstrapping and prompting strategies to alleviate the position bias and popularity bias issues that LLM-based ranking models may suffer. 

Extensive empirical studies indicate that LLMs have  promising zero-shot ranking abilities.
The empirical studies demonstrate the strong potential of transferring knowledge from LLMs as powerful recommendation models.
We aim at shedding light on several promising directions to further improve the ranking abilities of LLMs, including
(1) better perceiving the order of sequential historical interactions
and (2) alleviating the position bias and popularity bias.
For future work, we consider developing  technical approaches  to solve the above-mentioned key challenges when deploying LLMs as recommendation models. 
We also would like to develop LLM-based recommendation models that can be efficiently tuned on downstream user behaviors for effective personalized recommendations.

\section{Limitations}

In most experiments in this paper, ChatGPT is used as the primary target LLM for evaluation. However, being a closed-source commercial service, ChatGPT might integrate additional techniques with its core large language model to improve performance. While there are open-source LLMs available, such as LLaMA 2~\cite{touvron2023llama2} and Mistral~\cite{jiang2023mistral}, they exhibit a notable performance disparity compared to ChatGPT (e.g., LLaMA-2-70B-Chat vs. ChatGPT in Table~\ref{tab:llm_farm}). This gap makes it difficult to evaluate the emergent abilities of LLMs on the recommendation tasks using purely open-source models. In addition, we should note that the observations might be biased by specific prompts and datasets.

\section*{Acknowledgements}

This work was partially supported by National Natural Science Foundation of China under Grant No. 62222215, Beijing Natural Science Foundation under Grant No. L233008 and 4222027. Xin Zhao is the corresponding author.

\newpage

\bibliographystyle{splncs04}
\bibliography{reference}

\end{document}

%% file: sec-1-intro.tex
\section{Introduction}
\label{sec:intro}
In the literature of recommender systems, most existing models are trained with user behavior data from a specific domain or scenario~\cite{rendle2009bpr,he2020lightgcn,hidasi2016gru4rec}, suffering from two major issues.
Firstly, it is difficult to capture user preference by solely modeling historical behaviors, \eg  clicked item sequences~\cite{hidasi2016gru4rec,kang2018sasrec,zhou2020s3rec}, limiting the expressive power to model more complicated but explicit user interests (\eg intentions expressed in natural language).
Secondly, these models are essentially ``\emph{narrow experts}'',  lacking  more comprehensive knowledge in solving complicated recommendation tasks that rely on background or commonsense knowledge~\cite{guo2020survey_kg4rec_survey}. %

To improve  recommendation performance and interactivity, there have been increasing efforts that explore the use of pre-trained language models (PLMs) in recommender systems~\cite{geng2022p5,hou2022unisrec,wang2022unicrs}. 
They aim to explicitly capture  user preference in natural language~\cite{geng2022p5} or transfer rich world knowledge from text corpora~\cite{hou2022unisrec,hou2023vqrec}. 
Despite their effectiveness, thoroughly fine-tuning the recommendation models on task-specific data is still a necessity, making it less capable of solving diverse recommendation tasks~\cite{hou2022unisrec}.
More recently, large language models~(LLMs) 
have shown great potential to serve as zero-shot task solvers~\cite{wei2022flan,sanh2022t0}. Indeed, there are  some preliminary attempts that employ LLMs for solving  recommendation tasks~\cite{gao2023chat,wang2023zero,wang2023generative,dai2023uncovering,lin2023sparks,zhang2023instructrec}.  
These studies mainly focus on discussing  the possibility of building a capable recommender with LLMs.
While promising, the insufficient understanding of the new characteristics when making recommendations using LLMs could hinder the development of this new paradigm.

In this paper, we conduct empirical studies to investigate what determines the capacity of LLMs that serve as recommendation models. Typically, recommender systems are developed in a pipeline architecture~\cite{covington2016youtube}, consisting of \emph{candidate generation} ({retrieving relevant items}) and \emph{ranking} ({ranking relevant items at a higher position}) procedures.  %
{This work mainly focuses on the ranking stage of recommender systems, since LLMs are more expensive to run on a large-scale candidate set.  
Further, the ranking performance is sensitive to the retrieved candidate items, which is more suitable to examine the subtle differences in the recommendation abilities of LLMs.  
}

To carry out this study, we first formalize the recommendation process of LLMs as a \emph{conditional ranking} task. Given prompts that include sequential historical interactions as \emph{``conditions''}, LLMs are instructed to rank a set of \emph{``candidates''} (\eg items retrieved by candidate generation models), according to LLM's  intrinsic knowledge.
Then we conduct control experiments to systematically study the empirical performance of LLMs as rankers by  designing specific configurations for  ``conditions'' and ``candidates'', respectively. Overall, we attempt to answer the following key questions:
\begin{itemize}[leftmargin=0.8cm]
  \item What factors affect the zero-shot ranking performance of LLMs?
  \item What data or knowledge do LLMs rely on for recommendation?
\end{itemize}

Our empirical experiments are conducted on two public datasets for recommender systems. 
The results lead to several key findings that potentially shed light on how to develop LLMs as powerful ranking models for recommender systems.
We summarize the key findings as follows: 
\begin{itemize}[leftmargin=0.8cm]
  \item LLMs \textit{struggle to perceive the order} of the given sequential interaction histories. By employing specifically designed promptings, \textit{LLMs can be triggered to perceive the order}, leading to  improved ranking performance. 
  \item LLMs suffer from position bias and popularity bias while ranking, which can be alleviated by bootstrapping or specially designed prompting strategies.
  \item LLMs outperform existing zero-shot recommendation methods, showing promising zero-shot ranking abilities, especially on candidates retrieved by multiple candidate generation models with different practical strategies.
\end{itemize}

%% file: sec-2-method.tex
\section{General Framework for LLMs as Rankers}

To investigate the recommendation abilities of LLMs, we first formalize the recommendation process as a conditional ranking task. Then, we describe a general framework that adapts LLMs to solve the recommendation task.

\subsection{Problem Formulation} \label{sec:task}

Given the historical interactions $\mathcal{H}=\{i_{1}, i_{2},\ldots, i_{n}\}$ of one user (in chronological order of interaction time) as \emph{conditions}, the task is to rank the \emph{candidate} items $\mathcal{C} = \{i_{j}\}_{j=1}^{m}$, such that the items of interest would be ranked at a higher position.  In practice,
the candidate items are usually retrieved by candidate generation models from the whole item set $\mathcal{I}$  ($m \ll |\mathcal{I}|$)~\cite{covington2016youtube}. Further, we assume that each item $i$ is associated with a descriptive text $t_i$ following \cite{hou2022unisrec}.

\begin{figure}[!t]
    \centering
    \subfigure{\includegraphics[width=\textwidth]{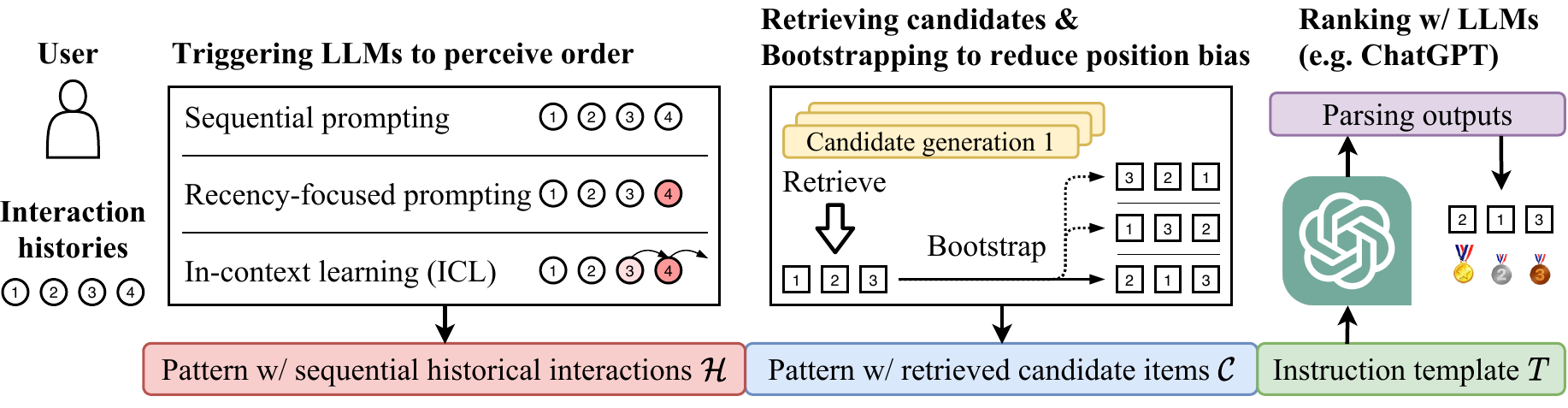}}
    \caption{An overview of the proposed LLM-based zero-shot ranking method.}
    \label{fig:model}
\end{figure}

\subsection{Ranking with LLMs Using Natural Language Instructions} \label{sec:ranking}

We use LLMs as ranking models to solve the above-mentioned task in an instruction-following paradigm~\cite{wei2022flan}. Specifically, for each user, we first construct two natural language patterns that contain sequential interaction histories $\mathcal{H}$  (\emph{conditions}) and retrieved candidate items $\mathcal{C}$ (\emph{candidates}), respectively. 
Then these patterns are filled into a natural language template $T$ as the final instruction. In this way, LLMs are expected to understand the instructions and output the ranking results as the instruction suggests. The overall framework of the ranking approach by LLMs is depicted in Figure~\ref{fig:model}. Next, we describe the detailed  instruction design  in our approach. 

\paratitle{Sequential historical interactions.} To investigate whether LLMs can capture user preferences from historical user behaviors, we include sequential historical interactions $\mathcal{H}$ into the instructions as inputs of LLMs. To enable LLMs to be aware of the sequential nature of historical interactions, we propose  three ways to construct the instructions:
\begin{itemize}[leftmargin=0.8cm]
    \item \textbf{Sequential prompting}:  Arrange the historical interactions in  chronological order.
    This way has also been used in prior studies~\cite{dai2023uncovering}.  
    For example, \emph{``I've watched the following movies in the past in order: '0. Multiplicity', '1. Jurassic Park', $\ldots$''}.
    \item \textbf{Recency-focused prompting}: In addition to the sequential interaction records, we can add an additional sentence to emphasize the most recent interaction. For example, \emph{``I've watched the following movies in the past in order: '0. Multiplicity', '1. Jurassic Park', $\ldots$. \candidate{Note that my most recently watched movie is Dead Presidents.} $\ldots$''}.
    \item \textbf{In-context learning~(ICL)}: %
    ICL is a prominent prompting approach for LLMs to solve various  tasks~\cite{zhao2023llm_survey}, where it includes demonstration examples in the prompt. 
    For the personalized recommendation task, simply introducing examples of other users may introduce noises because users usually have different preferences.
    Instead, we introduce demonstration examples by augmenting the input interaction sequence itself.
    We pair the prefix of the input interaction sequence and the corresponding successor as examples.
    For instance, \emph{``\candidate{If} I've watched the following movies in the past in order: '0. Multiplicity', '1. Jurassic Park', $\ldots$, \candidate{then you should recommend Dead Presidents to me and now that I've watched Dead Presidents, then} $\ldots$''}.
\end{itemize}

\paratitle{Retrieved candidate items.} Typically, candidate items to be ranked are first retrieved by candidate generation models~\cite{covington2016youtube}. 
In this work, we consider a relatively small pool for the candidates, and keep $20$ candidate items (\ie $m=20$) for ranking.
To rank these candidates with LLMs, we arrange the candidate items $\mathcal{C}$ 
in a sequential manner. For example, \emph{``Now there are 20 candidate movies that I can watch next: '0. Sister Act', '1. Sunset Blvd', $\ldots$''}. Note that,  following the classic candidate generation approach~\cite{covington2016youtube}, there is no specific order for candidate items. 
As a result,
We generate different  orders for the candidate items in the prompts, which enables us to further examine whether the ranking results of LLMs are affected by the arrangement order of candidates, \ie position bias, and how to alleviate position bias via bootstrapping. 

\paratitle{Ranking with large language models.} 
Existing studies show that LLMs can follow natural language instructions to solve diverse tasks in a zero-shot setting~\cite{wei2022flan,zhao2023llm_survey}. To rank using LLMs,
 we infill the patterns above into the instruction template $T$. An example  instruction template can be given as: \emph{``\condition{[pattern that contains sequential historical interactions $\mathcal{H}$]} \condition{[pattern that contains retrieved candidate items $\mathcal{C}$]} Please rank these movies by measuring the possibilities that I would like to watch next most, according to my watching history.''}.

\paratitle{Parsing the output of LLMs.} 
Note that the output of LLMs is still in natural language text, and we parse the output with heuristic text-matching methods and ground the recommendation results on the specified  item set. In detail, 
we can directly perform efficient substring matching algorithms like KMP~\cite{knuth1977fast_kmp} between the LLM outputs and the text of candidate items.
We also found that LLMs occasionally generate items that are not present in the candidate set. For GPT-3.5, such deviations occur in a mere 3\% of cases.
One can either reprocess the illegal cases or simply treat the out-of-candidate items as incorrect recommendations.

%% file: sec-3-experiment.tex
\begin{table}[!t]
	\setlength\tabcolsep{5pt}
        \small
	\centering
	\caption{Statistics of the preprocessed datasets. ``Avg. $|\mathcal{H}|$'' denotes the average length of historical interactions. ``Avg. $|t_i|$'' denotes the average number of tokens in the item text.}
    \label{tab:dataset}
	\begin{tabular}{crrrrrr}
    \toprule
    Dataset & \#Users & \#Items & \#Interactions & Sparsity & Avg. $|\mathcal{H}|$ & Avg. $|t_i|$ \\
    \midrule
    ML-1M & 6,040 & 3,706 & 1,000,209 & 95.53\% & 46.19 & 16.96 \\
    Games & 50,547 & 16,859 & 389,718 & 99.95\% & 7.02 & 43.31 \\
    \bottomrule
    \end{tabular}
\end{table}

\section{Empirical Studies}
\label{sec:empirical_studies}

\paratitle{Datasets.} The experiments are conducted on two widely-used public datasets for recommender systems: (1) the movie rating dataset \emph{MovieLens-1M}~\cite{harper2015movielens} (in short, \textbf{ML-1M}) where user ratings are regarded as interactions, and (2) one category from the \emph{Amazon Review} dataset~\cite{ni2019justifying} named \textbf{Games} where reviews are regarded as interactions. We filter out users and items with fewer than five interactions. Then we sort the interactions of each user by timestamp, with the oldest interactions first,
to construct the corresponding historical interaction sequences. The movie/product titles are used as the descriptive text of an item. We use item titles in this study for two reasons: (1) to determine if LLMs can make recommendations based on their intrinsic world knowledge with minimal information provided, and (2) to conserve computational resources. Exploring how LLMs use more extensive textual features for recommendations will be the focus of our future work.

\paratitle{Evaluation and implementation details.} Following existing works~\cite{kang2018sasrec,hou2022unisrec}, we apply the leave-one-out strategy for evaluation. For each historical interaction sequence, the last item is used as the ground-truth item in test set. The item before the last one is used in the validation set (used for training baseline methods). We adopt the widely used metric NDCG@K (in short, N@K) to evaluate the ranking results over the given $m$ candidates, where $K \le m$.  To ease the reproduction of this work, our experiments are conducted using a popular open-source recommendation  library \textsc{RecBole}~\cite{zhao2021recbole}. The historical interaction  sequences are truncated within a length of $50$. We evaluate LLM-based methods on all users in ML-1M dataset and randomly sampled $6,000$ users for Games dataset by default.
Unless specified, the evaluated LLM is accessed by calling OpenAI's API \texttt{gpt-3.5-turbo}.
The hyperparameter temperature of calling LLMs is set to $0.2$. \emph{All the reported results are the average of at least three repeat runs to reduce the effect of randomness.}

\subsection{Can LLMs Understand Prompts that Involve Sequential Historical User Behaviors?}\label{sec:his_order}

In LLM-based methods, historical interactions are naturally arranged in an ordered sequence.
By designing different configurations of $\mathcal{H}$, we aim to examine whether 
LLMs can leverage these historical user behaviors and perceive the sequential nature for making accurate recommendations.

\begin{figure}[!t]
    \centering
    \subfigure[Perceive Order]{\label{fig:his:order}\includegraphics[width=0.30\textwidth]{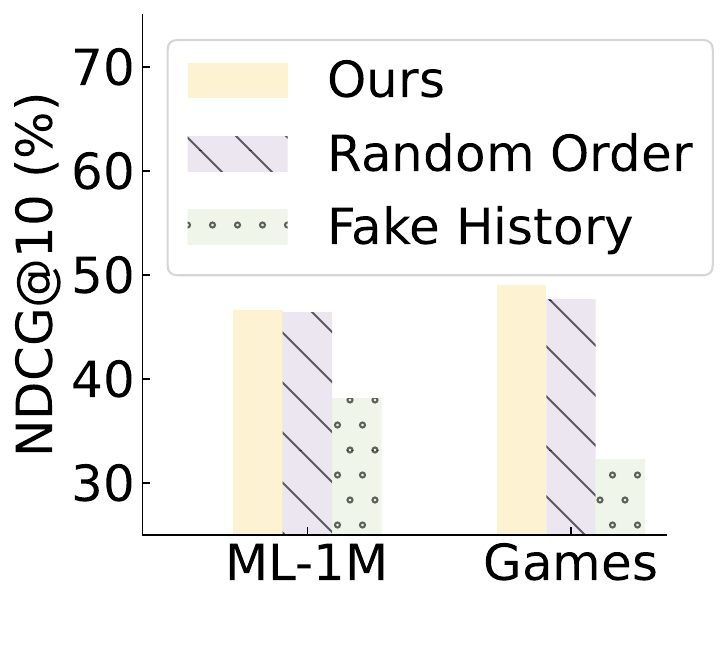}}
    \hspace{25px}
    \subfigure[Performance vs. $|\mathcal{H}|$]{\label{fig:his:len}\includegraphics[width=0.28\textwidth]{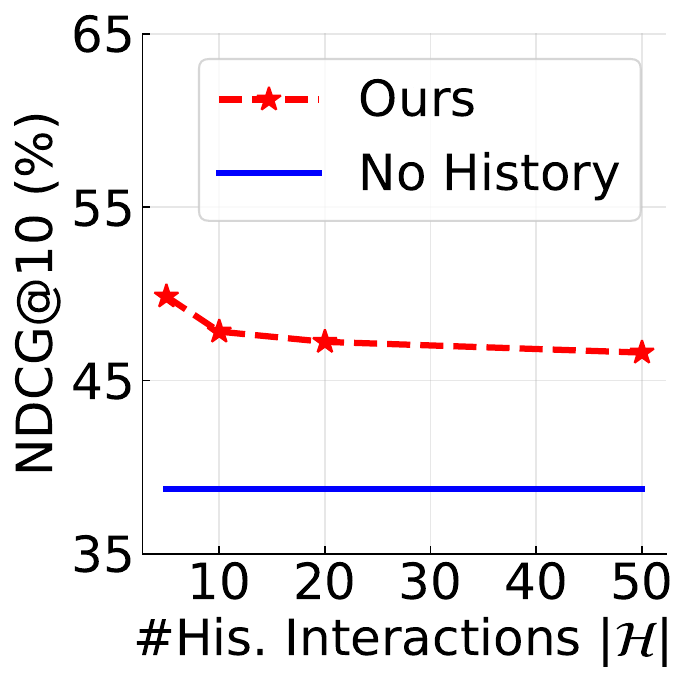}}
    \caption{Analysis of whether LLMs perceive the order of historical interactions.}
    \label{fig:his}
\end{figure}

\paratitle{LLMs struggle to perceive the order of given historical user behaviors.} In this section, we examine whether LLMs can understand prompts with ordered historical interactions and give personalized recommendations. The
task is to rank a candidate set of $20$ items, containing one ground-truth item
and $19$ randomly sampled negatives.
By analyzing historical behaviors, items of interest should be
ranked at a higher position. We compare the ranking results of three LLM-based methods: (a) \emph{Ours}, which ranks as we have described 
in Section~\ref{sec:ranking}. Historical user behaviors are encoded into prompts using the ``sequential prompting'' strategy. 
(b) \emph{Random Order}, where the historical user behaviors will be randomly shuffled before being fed to the model, and 
(c) \emph{Fake History}, where we replace all the items in original historical behaviors with randomly sampled items as fake historical behaviors.
From Figure~\ref{fig:his:order}, we can see that \emph{Ours} has better performance than variants with fake historical behaviors. 
However, the performance of \emph{Ours} and \emph{Random Order} is similar,  indicating that LLMs are not sensitive to the order of the given historical user interactions.

Moreover, in Figure~\ref{fig:his:len}, we vary the number of latest historical user behaviors ($|\mathcal{H}|$) used for constructing the prompt  from $5$ to $50$. The results show that 
increasing the number of historical user behaviors does not improve, but rather negatively impacts the ranking performance. We speculate that this phenomenon is caused by the fact that \textcolor{black}{LLMs have difficulty understanding the order, but consider all the historical behaviors equally. Therefore too many historical user behaviors (\eg $|\mathcal{H}|=50$) may overwhelm LLMs and lead to a performance drop.} In contrast, a relatively small $|\mathcal{H}|$ enables LLMs to concentrate on the most recently interacted items, resulting in better recommendation performance.

\paratitle{Triggering LLMs to perceive the interaction order.} Based on the above observations, we find it difficult for LLMs to perceive the order in interaction histories by a default prompting strategy.
As a result, 
we aim to elicit the order-perceiving abilities of LLMs, by proposing two alternative prompting strategies and emphasizing the recently interacted items.
Detailed descriptions of the proposed strategies have been given in Section~\ref{sec:ranking}.
In Table~\ref{tab:seq}, we can see that both recency-focused prompting and in-context learning can generally improve the ranking performance of LLMs, though the best strategy may vary on different datasets. 
The above results can be summarized as the following key observation:
\begin{mdframed}
\textbf{Observation 1.} LLMs \textit{struggle to perceive the order} of the given sequential interaction histories. By employing specifically designed promptings, \textit{LLMs can be triggered to perceive the order} of historical user behaviors, leading to  improved ranking performance. 
\end{mdframed}

\begin{table}[!t]
	\setlength\tabcolsep{3pt}
	\centering
        \small
	\caption{Performance comparison on \emph{randomly retrieved candidates}. Ground-truth items are included in the candidate sets. 
 \textcolor{black}{``full'' denotes models that are trained on the target dataset, and ``zero-shot'' denotes models that are not trained on the target dataset but could be pre-trained.
 We highlight the best performance among zero-shot recommendation methods in \textbf{bold}.} }
	\begin{tabular}{ccrrrrrrrr}
	\toprule
	& \multirow{2.5}[0]{*}{Method} & \multicolumn{4}{c}{ML-1M} & \multicolumn{4}{c}{Games} \\
	\cmidrule(lr){3-6} \cmidrule(lr){7-10}
	& & N@1  & N@5    & N@10 & N@20 & N@1  & N@5    & N@10 & N@20\\
	\midrule
        \parbox[t]{3mm}{\multirow{3}{*}{\rotatebox[origin=c]{90}{\small full}}}
	& Pop & 22.91 & 45.16 & 52.33 & 55.36 & 28.35 & 47.42&52.96 	&57.45  \\
        & BPRMF~\cite{rendle2009bpr} & 34.60 & 59.87 & 64.29 & 65.39 & 44.92 & 62.33 & 66.27 & 68.94 \\
        & SASRec~\cite{kang2018sasrec} & 61.39 & 76.39 & 78.89 & 79.79 & 56.90 & 73.19 & 75.92 & 77.14   \\
	\midrule
	\parbox[t]{3mm}{\multirow{6.5}{*}{\rotatebox[origin=c]{90}{\small zero-shot}}}
        & BM25~\cite{robertson2009bm25} &  4.70 & 12.68 & 17.88 & 33.19 & 13.92 & 28.81 & 34.61 & 44.35   \\
        & UniSRec~\cite{hou2022unisrec} & 7.37 & 18.80 & 26.67 & 37.93 & 18.95 & 33.99 & 40.71 & 48.42  \\
        & VQ-Rec~\cite{hou2023vqrec} & 5.98 & 15.48 & 23.74 & 35.85 & 7.28 & 18.28 & 26.21 & 37.62\\
        \cmidrule(lr){2-10}
	& Sequential & 18.28 & 36.35 & 42.85 & 49.02 & 30.28 & 45.48 & 50.57 & 56.55    \\
        & Recency-Focused & 19.57 & 37.73 & 44.23 & 50.01 & \textbf{34.03} & \textbf{48.77} & \textbf{53.50} & \textbf{59.01}    \\
        & In-Context Learning & \textbf{21.77} & \textbf{39.59} & \textbf{45.83} & \textbf{51.62} & 33.95 & 48.44 & 53.10 & 58.92 \\
	\bottomrule
	\end{tabular}
	\label{tab:seq}
\end{table}

\subsection{Do LLMs suffer from biases while ranking?}\label{sec:bias}

The biases and debiasing methods in conventional recommender systems have been widely studied~\cite{chen2020bias}. 
For LLM-based recommendation models, both the input and output are natural language texts and will inevitably introduce new biases.
In this section, we discuss two kinds of biases that LLM-based recommendation models suffer from. We also make discussions on how to alleviate these biases. 

\begin{figure}[!t]
    \centering
    \subfigure[Position Bias]{\label{fig:bias:pos}\includegraphics[width=0.24\textwidth]{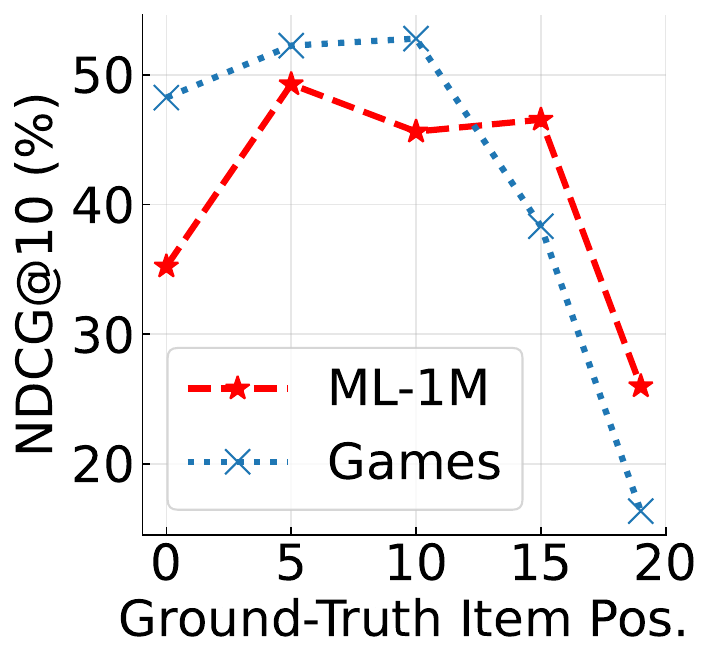}}
    \subfigure[Bootstrapping]{\label{fig:bias:boot}\includegraphics[width=0.24\textwidth]{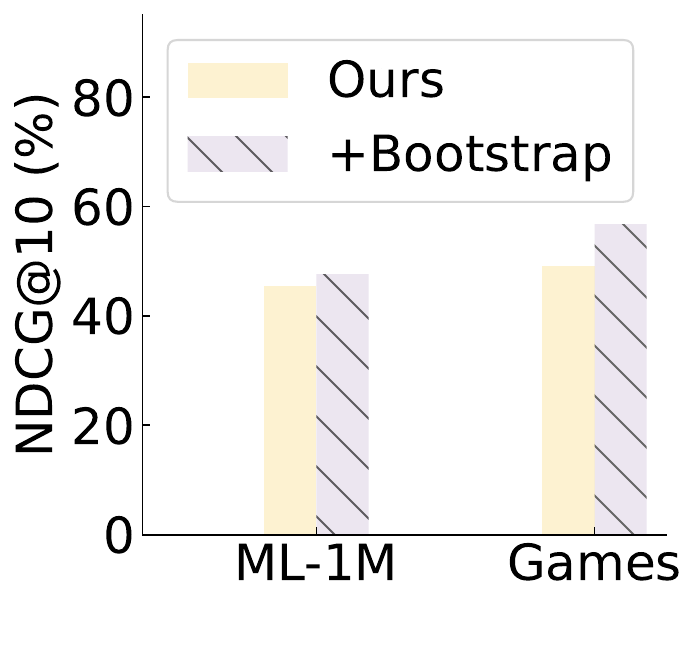}}
    \subfigure[Popularity Bias]{\label{fig:bias:pop}\includegraphics[width=0.24\textwidth]{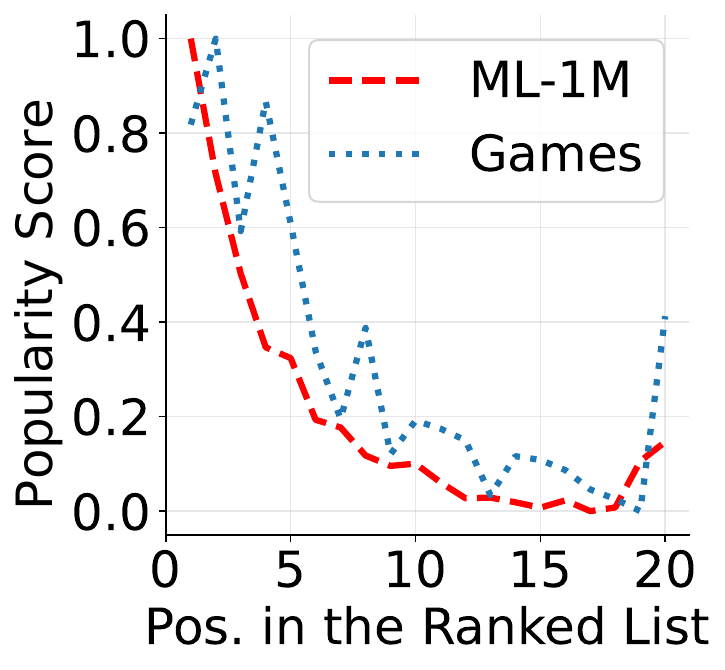}}
    \subfigure[Popularity of top-$1$ item \emph{w.r.t.} $|\mathcal{H}|$]{\label{fig:bias:len_pop}\includegraphics[width=0.24\textwidth]{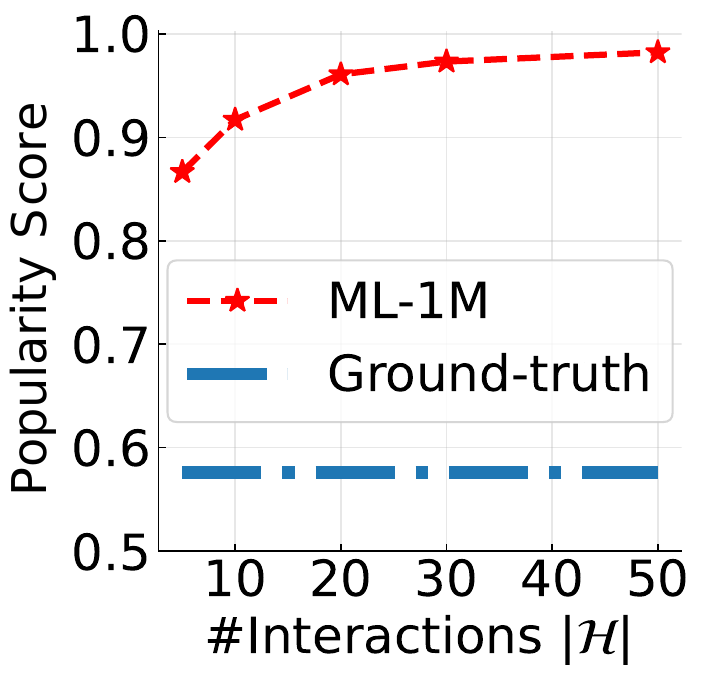}}
    \caption{Biases and debiasing methods in the ranking of LLMs. (a) The position of candidates in the prompts influences the ranking results. (b) Bootstrapping alleviates position bias. (c) LLMs tend to recommend popular items. (d) Focusing on historical interactions reduces popularity bias.}
    \label{fig:bias}
\end{figure}

\paratitle{The order of candidates affects the ranking results of LLMs.} For conventional ranking methods, the order of retrieved candidates usually will not affect the ranking results~\cite{kang2018sasrec,hidasi2016gru4rec}.
However, for the LLM-based approach that is described in Section~\ref{sec:ranking}, the candidates are arranged in a sequential manner and infilled into a prompt.
It has been shown that LLMs are generally sensitive to the order of examples in the prompts for NLP tasks~\cite{zhao2021calibrate,lu2022fantastically}.
As a result, we also conduct experiments to examine whether the order of candidates affects the ranking performance of LLMs.
We follow the experimental settings adopted in Section~\ref{sec:his_order}.
The only difference is that we control the order of these candidates in the prompts, by making the ground-truth items appear at a certain position.
We vary the position of ground-truth items at $\{0, 5, 10, 15, 19\}$ and present the results in Figure~\ref{fig:bias:pos}.
We can see that the performance varies when the ground-truth items appear at different positions. Especially, the ranking performance drops significantly when the ground-truth items appear at the last few positions.
The results indicate that LLM-based rankers are affected by the order of candidates, \ie \emph{position bias}, which may \emph{not} affect conventional recommendation models.

\paratitle{Alleviating position bias via bootstrapping.} 
A simple strategy to alleviate position bias is to bootstrap the ranking process. We may rank the candidate set repeatedly for $B$ times, with candidates randomly shuffled at each round. In this way, one candidate may appear in different positions.
We then merge the results of each round to derive the final ranking. From Figure~\ref{fig:bias:boot}, 
we follow the setting in Section~\ref{sec:his_order} and apply the bootstrapping strategy to \emph{Ours}. Each candidate set will be ranked for $3$ times. We can see that bootstrapping improves the ranking performance on both datasets.

\paratitle{Popularity degrees of candidates affect ranking results of LLMs.} For  popular items, the associated text may also appear frequently in the pre-training corpora of LLMs. For example,  a best-selling book would be widely discussed on the Web. Thus, we aim to examine whether the ranking results are affected by  the popularity of candidates. However, it is difficult to directly measure the popularity of item text. Here, we hypothesize that the text popularity can be indirectly measured by item frequency in one recommendation dataset. In Figure~\ref{fig:bias:pop}, we report the item popularity score (measured by the normalized item frequency of appearance in the training set) at each position of the ranked item lists. We can see that popular items tend to be ranked at higher positions. 

\paratitle{Making LLMs focus on historical interactions helps reduce popularity bias.}
We assume that if LLMs focus on historical interactions, they may give more personalized recommendations but not more popular ones. 
From Figure~\ref{fig:his:len}, we know that LLMs make better use of historical interactions when using less historical interactions.
From Figure~\ref{fig:bias:len_pop}, we compare the popularity scores of the best-ranked items varying the number of historical interactions.
It can be observed that as $|\mathcal{H}|$ decreases, the popularity score decreases as well. This suggests that one can reduce the effects of popularity bias when LLMs focus more on historical interactions.
From the above experiments, we can conclude the following:
\begin{mdframed}
\textbf{Observation 2.} LLMs suffer from position bias and popularity bias while ranking, which can be alleviated by bootstrapping or specially designed prompting strategies.
\end{mdframed}

\subsection{How Well Can LLMs Rank Candidates in a Zero-Shot Setting?} \label{sec:candidate}

We further evaluate LLM-based methods on candidates with hard negatives that are retrieved by different strategies to further investigate what the ranking of LLMs depends on.
Then, we present the ranking performance of different methods on candidates retrieved by multiple candidate generation models to simulate a more practical and difficult setting.

\begin{table}[!t]
	\setlength\tabcolsep{2pt}
	\centering
        \small
	\caption{Zero-shot ranking performance comparison. 
 We highlight the best performance in \textbf{bold}. Due to limited budget, we evaluate each LLM only once on $200$ sampled users \textbf{only} for experiments corresponding to this table.}
	\begin{tabular}{crrrrrrrr}
	\toprule
	\multirow{2.5}[0]{*}{Method} & \multicolumn{4}{c}{ML-1M} & \multicolumn{4}{c}{Games} \\
	\cmidrule(lr){2-5} \cmidrule(lr){6-9}
	& N@1  & N@5    & N@10 & N@20 & N@1  & N@5    & N@10 & N@20\\
	\midrule
        BM25~\cite{robertson2009bm25} &  4.70 & 12.68 & 17.88 & 33.19 & 13.92 & 28.81 & 34.61 & 44.35   \\
        UniSRec~\cite{hou2022unisrec} & 7.37 & 18.80 & 26.67 & 37.93 & 18.95 & 33.99 & 40.71 & 48.42  \\
        \midrule
		Alpaca-7B~\cite{taori2023stanford} & 4.00 & 13.92 & 23.09 & 31.54 & 5.50 & 14.16 & 21.67 & 28.68 \\
		Vicuna-13B~\cite{chiang2023vicuna} & 6.50 & 14.75 & 22.64 & 33.42 & 7.00 & 17.73 & 24.30 & 31.22 \\
		LLaMA-2-70B-Chat~\cite{touvron2023llama2} & 8.00 & 25.42 & 31.19 & 34.52 & 21.50 & 32.30 & 37.83 & 41.97 \\
		ChatGPT (GPT-3.5) & \textbf{23.33} & \textbf{42.07} & \textbf{48.80} & \textbf{53.73} & 23.83 & 45.69 & 50.31 & 55.45 \\
		GPT-4 & 15.50 & 40.65 & 46.74 & 48.42 & \textbf{39.50} & \textbf{58.22} & \textbf{62.88} & \textbf{65.25} \\
	\bottomrule
	\end{tabular}
	\label{tab:llm_farm}
\end{table}

\paratitle{LLMs have promising zero-shot ranking abilities.} In Table~\ref{tab:seq}, we conduct experiments to compare the ranking abilities of LLM-based methods with existing methods. We follow the same setting in Section~\ref{sec:his_order} where $|\mathcal{C}| = 20$ and candidate items are randomly retrieved. We include three conventional models that are trained on the training set, \ie Pop (recommending according to item popularity), BPRMF~\cite{rendle2009bpr}, and SASRec~\cite{kang2018sasrec}. We also evaluate three zero-shot recommendation methods that are not trained on the target datasets, including BM25~\cite{robertson2009bm25} (rank according to the textual similarity between candidates and historical interactions), UniSRec~\cite{hou2022unisrec}, and VQ-Rec~\cite{hou2023vqrec}. For UniSRec and VQ-Rec, we use their publicly available pre-trained models. We do not include ZESRec~\cite{ding2021zero} because there is no pre-trained model released.
In addition, we compare the zero-shot ranking performance of different LLMs in Table~\ref{tab:llm_farm}. ``Recency-Focused'' prompting strategy is used for LLM-based rankers.

From Table~\ref{tab:seq} and~\ref{tab:llm_farm}, we can see that LLMs with more parameters generally perform better. The best LLM-based methods outperform existing zero-shot recommendation methods by a large margin, showing promising zero-shot ranking abilities. We would highlight that it is difficult to conduct zero-shot recommendations on the ML-1M dataset, due to the difficulty in measuring the similarity between movies merely by the similarity of their titles.
However, LLMs can use their intrinsic knowledge to measure the similarity between movies and make recommendations.
We would emphasize that the goal of evaluating zero-shot recommendation methods is not to surpass conventional models. The goal is to demonstrate the strong recommendation capabilities of pre-trained base models, which can be further adapted and transferred to downstream scenarios.

\begin{figure}[!t]
    \centering
    \subfigure{\includegraphics[width=\textwidth]{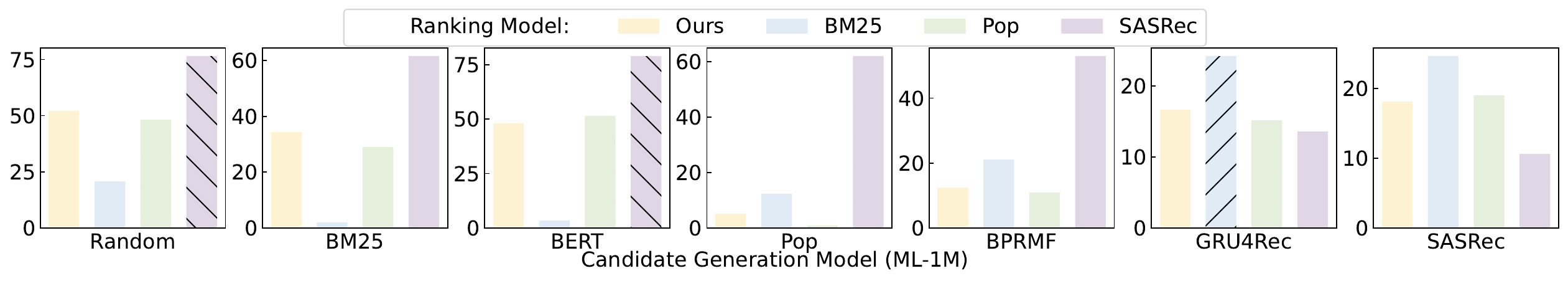}}
    \subfigure{\includegraphics[width=\textwidth]{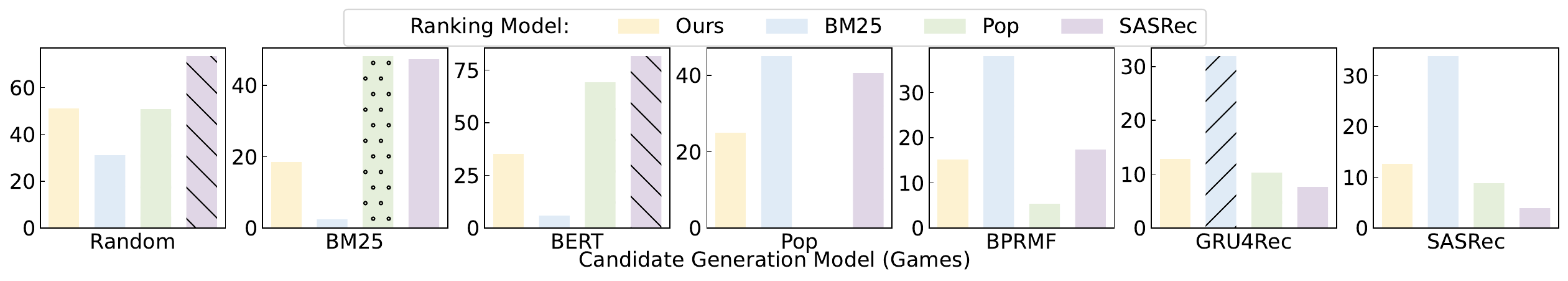}}
    \caption{Ranking performance measured by NDCG@10 (\%) on hard negatives.}
    \label{fig:source}
\end{figure}

\paratitle{LLMs rank candidates based on item popularity, text features as well as user behaviors.}
To further investigate how LLMs rank the given candidates, we evaluate LLMs on candidates that are retrieved by different candidate generation methods.
These candidates can be viewed as hard negatives for ground-truth items, which can be used to measure the ranking ability of LLMs for specific categories of items.
We consider two categories of strategies to retrieve the candidates:
(1) \emph{content-based methods} like \emph{BM25}~\cite{robertson2009bm25} and \emph{BERT}~\cite{devlin2019bert} retrieve candidates based on the text feature similarities, and
(2) \emph{interaction-based methods}, including \emph{Pop}, \emph{BPRMF}~\cite{rendle2009bpr}, \emph{GRU4Rec}~\cite{hidasi2016gru4rec}, and \emph{SASRec}~\cite{kang2018sasrec}, retrieve items using neural networks trained on user-item interactions.
Given candidates, we compare the ranking performance of the LLM-based model (\emph{Ours}) and representative methods.

From Figure~\ref{fig:source}, we can see that the ranking performance of the LLM-based method varies on different candidate sets and different datasets. (1) On ML-1M, LLM-based method cannot rank well on candidate sets that contain popular items (\eg \emph{Pop} and \emph{BPRMF}), indicating the LLM-based method recommend items largely depend on item popularity on ML-1M dataset.  (2) On Games, we can observe that \emph{Ours} has similar performance both on popular candidates and textual similar candidates, showing that item popularity and text features contribute similarly to the ranking of LLMs. (3) On both two datasets, the performance of \emph{Ours} is affected by hard negatives retrieved by interaction-based candidate generation models, but not as severe as those interaction-based rankers like \emph{SASRec}. The above results demonstrate that LLM-based methods not only consider one single aspect for ranking, but make use of item popularity, text features, and even user behaviors. On different datasets, the weights of these three aspects to affect the ranking performance may also vary.

\begin{table}[!t]
	\setlength\tabcolsep{4pt}
	\centering
        \small
	\caption{Performance comparison on \emph{candidates retrieved by multiple candidate generation models}. Ground-truth items are \emph{not} guaranteed to be included in the candidate sets. 
 ``full'' denotes models that are trained on the target dataset, and ``zero-shot'' denotes models that are not trained on the target dataset but could be pre-trained.
 We highlight the best and second-best performance among \emph{all} recommendation methods in \textbf{bold}.}
	\begin{tabular}{ccrrrrrrrr}
	\toprule
	& \multirow{2.5}[0]{*}{Method} & \multicolumn{4}{c}{ML-1M} & \multicolumn{4}{c}{Games}\\
	\cmidrule(lr){3-6} \cmidrule(lr){7-10} 
	& & N@1  & N@5    & N@10 & N@20 & N@1  & N@5    & N@10 & N@20\\
	\midrule
        \parbox[t]{3mm}{\multirow{3}{*}{\rotatebox[origin=c]{90}{\small full}}}
	& Pop & 0.08 & 1.20 & 4.13 & 5.79 & 0.13 & 1.00 & 2.27 & 2.62\\
        & BPRMF~\cite{rendle2009bpr}  & 0.26 & 1.69 & 4.41 & 6.04 & 0.55 & 1.98 & \textbf{2.96} & \textbf{3.19} \\
        & SASRec~\cite{kang2018sasrec}  & \textbf{3.76} & \textbf{9.79} & \textbf{10.45} & \textbf{10.56} & \textbf{1.33} & \textbf{3.55} & \textbf{4.02} & \textbf{4.11}   \\
	\midrule
	\parbox[t]{3mm}{\multirow{4.5}{*}{\rotatebox[origin=c]{90}{\small zero-shot}}}
        & BM25~\cite{robertson2009bm25} & 0.26 & 0.87 & 2.32 & 5.28 & 0.18 & 1.07 & 1.80 & 2.55\\
        & UniSRec~\cite{hou2022unisrec} &0.88 & 3.46 & 5.30 & 6.92 & 0.00 & 1.86 & 2.03 & 2.31 \\
        & VQ-Rec~\cite{hou2023vqrec} &0.20 & 1.60 & 3.29 & 5.73 & 0.20 & 1.21 & 1.91 & 2.64\\
        \cmidrule(lr){2-10}
	& Ours   & \textbf{1.74} & \textbf{5.22} & \textbf{6.91} & \textbf{7.90} & \textbf{0.90} & \textbf{2.26} & 2.80 & 3.08\\
	\bottomrule
	\end{tabular}
	\label{tab:multi}
\end{table}

\paratitle{LLMs can effectively rank candidates retrieved by multiple candidate generation models.} For real-world recommender systems~\cite{covington2016youtube}, the items to be ranked are usually retrieved by multiple candidate generation models. As a result, we also conduct experiments in a more practical and difficult setting. We use the above-mentioned seven candidate generation models to retrieve items.
The top-$3$ best items retrieved by each candidate generation model will be merged into a candidate set containing a total of $21$ items. As a more practical setting, we do not complement the ground-truth item to each candidate set.
Note that the experiments here were conducted under the implicit preference setup~\cite{zhao2022revisiting}, indicating that implicit positive instances (not explicitly labeled) may exist among the retrieved items. A more faithful evaluation might require a human study, which we intend to explore in our future work.
For \emph{Ours}, we summarize the experiences gained from Section~\ref{sec:his_order} and~\ref{sec:bias}.
We use the recency-focused prompting strategy to encode $|\mathcal{H}| = 5$ sequential historical interactions into prompts and use a bootstrapping strategy to repeatedly rank for $3$ rounds.

From Table~\ref{tab:multi}, we can see that the LLM-based model (\emph{Ours}) yields the second-best performance over the compared recommendation models on most metrics.
The results show that LLM-based zero-shot ranker even beats the conventional recommendation model \emph{Pop} and \emph{BPRMF} that has been trained on the target datasets, further demonstrating the strong zero-shot ranking ability of LLMs.
We assume that LLMs can make use of their intrinsic world knowledge to 
rank the candidates comprehensively considering popularity, text features, and user behaviors. 
In comparison, existing models (as \emph{narrrow experts}) may lack the ability to rank items in a complicated setting.
The above findings can be summarized as:
\begin{mdframed}
\textbf{Observation 3.} LLMs have promising zero-shot ranking abilities, especially on candidates retrieved by multiple candidate generation models with different practical strategies.
\end{mdframed}

%% file: sec-4-related-work.tex
\section{Related Work}

\paratitle{Transfer learning for recommender systems.} 
As recommender systems are mostly trained on data collected from a single source, people have sought to transfer knowledge from other domains~\cite{zang-2023-cross-survey,zhu-2021-ijcai-cross-survey,man-2017-ijcai-cross,zhu-2022-wsdm-cross,zhao-2020-sigir-cross,zhu-2019-cikm-cross}, markets~\cite{bonab-2021-cikm-cross-market,roitero-2020-www-cross-market}, or platforms~\cite{cao-2017-cross-platforms,gao-2023-tkde-cross-platforms}.
Typical transfer learning methods for recommender systems rely on anchors, including shared users/items~\cite{man-2017-ijcai-cross,zhu-2020-ijcai-cross,yuan-2020-sigir-parameter_efficient,yuan-2021-sigir-one_person,chen-2023-user-specific,cheng-2021-icdm-learning} or representations from a shared space~\cite{cui-2020-recsys-cross,yuan-2023-exploring-adapter,yuan-2023-exploring-llm}.
However, these anchors are usually sparse among different scenarios, making transferring difficult for recommendations~\cite{zhu-2021-ijcai-cross-survey}.
More recently, there are studies aiming to transfer knowledge stored in language models by adapting them to recommendation tasks via tuning~\cite{bao2023tallrec_tuning_rec_hexiangnan,geng2022p5,cui2022m6_rec,Shin-2021-scaling-law} or prompting~\cite{li2023personalized_prompt_tuning_explainable_zhangyongfeng,zhang2023prompt_news_recommendation,li2023pbnr_prompt_news_recommendation}.
In this paper, we conduct zero-shot recommendation experiments to examine the potential to transfer knowledge from LLMs.

\paratitle{Large language models for recommender systems.} The design of recommendation models, especially sequential recommendation models, has been long inspired by the design of language models, from word2vec~\cite{barkan-2016-item2vec,airnb-2018-kdd-item_embedding,he17translation} to recent neural networks~\cite{hidasi2016gru4rec,kang2018sasrec,zhou2020s3rec,tang-2018-wsdm-caser}. In recent years, with the development of pre-trained language models (PLMs)~\cite{devlin2019bert}, people have tried to transfer knowledge stored in PLMs to recommendation models, by either representing items using their text features or representing behavior sequences in the format of natural language~\cite{geng2022p5,wang2022transrec_yuanfeijie,liu2023pre_plm_for_recommendation_survey,ZESRec,xiao-2022-kdd-news_plm}.
Very recently, large language models (LLMs) have been shown superior language understanding and generation abilities~\cite{zhao2023llm_survey,touvron2023llama,ouyang2022training_instruct_tuning_align_user_intent_instructgpt,wu2023survey_llm4rec,fan2023recommender,chen2023large,wu2023survey}. Studies have been made to make recommender systems more interactive by integrating LLMs along with conventional recommendation models~\cite{gao2023chat,liGPT4RecGenerativeFramework2023,liu2023generative_recommender_news,wang2023zero,he2023large,wang2023rethinking,wei2024llmrec,ren2023representation} or fine-tuned with specially designed instructions~\cite{cui2022m6_rec,geng2022p5,bao2023tallrec_tuning_rec_hexiangnan,hua2023index_generative_rec,zheng2023adapting}.
There are also early explorations showing LLMs have zero-shot recommendation abilities~\cite{wang2023zero,liuChatGPTGoodRecommender2023,dai2023uncovering,kang2023llms_evalution_rating_prediction,zhang2023chatgpt_fair_evaluation_rec,wang2023generative,wang2023recmind,zhang2023agentcf}.
Despite being effective to some extent, few works have explored what determines the recommendation performance of LLMs.